\documentclass[apj]{emulateapj}
\usepackage{color}
\usepackage{multirow}

\def\spose#1{\hbox to 0pt{#1\hss}}
\newcommand\lsim{\mathrel{\spose{\lower 3.0pt\hbox{$\mathchar"218$}}
     \raise 2.0pt\hbox{$\mathchar"13C$}}}
\newcommand\gsim{\mathrel{\spose{\lower 3.0pt\hbox{$\mathchar"218$}}
     \raise 2.0pt\hbox{$\mathchar"13E$}}}
\newcommand\msun{{\rm \,M_\odot}}

\usepackage{verbatim}
\usepackage{color}
\usepackage[usenames,dvipsnames]{xcolor}
\definecolor{green}{rgb}{0.0, 0.4, 0.0}
\definecolor{forestgreen(web)}{rgb}{0.13, 0.55, 0.13}
\definecolor{green(web)}{rgb}{0.13, 0.55, 0.13}
\definecolor{green}{rgb}{0.0, 0.4, 0.0}

\begin{document}

\title{Hydrodynamic Simulations of the Central Molecular Zone with a Realistic
Galactic Potential}
\author{Jihye Shin\altaffilmark{1}, Sungsoo S. Kim\altaffilmark{2,3},
        Junichi Baba\altaffilmark{4,5}, Takayuki R. Saitoh\altaffilmark{6},
        Jeong-Sun Hwang\altaffilmark{7},
        Kyungwon Chun\altaffilmark{3}, Shunsuke Hozumi\altaffilmark{8}}
\altaffiltext{1}{School of Physics, Korea Institute for Advanced Study, 85 Hoegiro, 
Dongdaemun-gu, Seoul 130-722, Korea}
\altaffiltext{2}{Department of Astronomy \& Space Science, Kyung Hee University,
Yongin, Gyeonggi 446-701, Korea}
\altaffiltext{3}{School of Space Research, Kyung Hee University, Yongin, Gyeonggi,
446-701, Korea}
\altaffiltext{4}{Research Center for Space and Cosmic Evolution, Ehime University
2-5 Bunkyo-cho, Matsuyama, Ehime, 790-8577, Japan}
\altaffiltext{5}{National Astronomical Observatory of Japan, Mitaka-shi, Tokyo 181-8588, Japan}
\altaffiltext{6}{Earth-Life Science Institute, Tokyo Institute of Technology,
2-12-1 Ookayama, Meguro-ku, Tokyo 152-8551, Japan}
\altaffiltext{7}{Department of Physics and Astronomy, Sejong University,
209 Neungdong-ro, Gwangjin-gu, Seoul 143-747, Korea}
\altaffiltext{8}{Faculty of Education, Shiga University, 2-5-1 Hiratsu, Otsu,
Shiga 520-0862, Japan}

\begin{abstract}
We present hydrodynamic simulations of gas clouds inflowing from the
disk to a few hundred parsec region of  the Milky Way. A gravitational potential is generated to include realistic Galactic structures by using thousands of multipole expansions that describe 6.4 million stellar particles of a self-consistent Galaxy simulation.
We find that a hybrid multipole expansion model, with two
different basis sets and a thick disk correction, accurately reproduces the overall
structures of the Milky Way. Through non-axisymmetric Galactic structures
of an elongated bar and spiral arms, gas clouds in the disk inflow to the
nuclear region and form a central molecular zone (CMZ)-like nuclear ring.
We find that the size of the nuclear ring evolves into $\sim240~$pc at 
$T\sim1500~$Myr, regardless of the initial size. For most simulation runs, the 
rate of gas inflow to the nuclear region is equilibrated to $\sim0.02\msun~\mathrm{yr^{-1}}$. The nuclear ring is off-centered, relative to the Galactic center, by the 
lopsided central mass distribution of the Galaxy model, and thus an asymmetric mass 
distribution of the nuclear ring arises accordingly. The vertical asymmetry of the the Galaxy model also causes the nuclear ring to be tilted along the Galactic plane. During the first $\sim100$~Myr, the vertical 
frequency of the gas motion is twice that of the orbital frequency, thus the projected nuclear 
ring shows a twisted, $\infty$-like shape. 
\end{abstract}

\section{Introduction}

\begin{table*}
\caption{Basis set for multipole expansion models\label{tbl-1}}
\centering
\begin{tabular}{ccl}
\tableline\tableline
ME model & Coordinate & Basis set\\
\tableline
\multirow{2}{*}{E96} &
\multirow{2}{*}{Cylindrical} &
$\rho_{kmh}(R,\phi,z) = \frac{1}{4\pi G}k^2~e^{-k|z-h|}e^{im\phi}~J_m(kR)$\\
&
&
$\Phi_{kmh}(R,\phi,z) = -\frac{1}{2}(1+k|z-h|)e^{-k|z-h|}~e^{im\phi}~J_m(kR)$\\
\tableline
\multirow{2}{*}{HO92} &
\multirow{2}{*}{Spherical} &
$\rho_{nlm}(r,\theta,\phi) = \frac{K_{nl}}{2\pi}\frac{r^l}{r(1+r)^{2l+3}}
                             \sqrt{4\pi}~Y_{lm}(\theta,\phi)~C_{n}^{(2l+3/2)}
                             \left(\frac{r-1}{r+1}\right)$\\
&
&
$\Phi_{nlm}(r,\theta,\phi) = -\frac{r^l}{(1+r)^{2l+1}}\sqrt{4\pi}~Y_{lm}
                             (\theta,\phi)~C_{n}^{(2l+3/2)}\left(\frac{r-1}
                             {r+1}\right)$\\
\tableline
\multirow{2}{*}{AI78} &
\multirow{2}{*}{Polar} &
$\mu_{nm}(R,\phi)  =\frac{Ma}{R_a^3}\frac{(2n+1)}{2\pi}\exp(im\phi)~P_{nm}(r)$\\
&
&
$\Psi_{nm}(R,\phi) =-\frac{GM}{R_a}\exp(im\phi)~P_{nm}(r)$ \\
\tableline
\end{tabular}
\tablecomments{$Y_{lm}(\theta, \phi)$, $C_n^{\alpha}(x)$, $J_{m}(x)$, and
$P_{nm}(x)$ are spherical harmonics, Gegenbauer polynomial, cylindrical
Bessel function, and Legendre function, respectively. Details are enumerated
in Appendices~A and B.}
\end{table*}
 
Emission from CO molecules reveals that the molecular gas in the Milky Way
is abundant along the Galactic plane down to a Galactocentric radius $R_g$
of $\sim3~$kpc \citep{dam01}. The amount of molecular gas is greatly reduced
inside the radius, and a mixed atomic/molecular layer named the ``HI nuclear disk''
appears instead \citep{mor96,fer07}. The molecular gas tends to become concentrated 
again inside $R_g \sim 200~$pc with a high number density of $n \gsim 10^4 \mathrm{cm}^{-3}$, 
and this region is referred to as the ``Central Molecular Zone'' (CMZ, \citealt{mor96}). 

Migration of gas clouds from the Galactic disk to the central few hundred parsecs is 
primarily induced by the rotating non-axisymmetric potential,
the Galactic bar \citep{bin91}. Gas clouds that are moving along the so-called
X$_1$ orbit, a family of stable, closed orbits elongated along the bar's major
axis, lose angular momentum near the orbit cusps and plunge to the
X$_2$ orbits inside, another a family of stable, closed orbits elongated along the bar's
minor axis in the deeper potential. The infalling gas clouds are
compressed when arriving at X$_2$ orbits by collision with resident gas clouds, 
and are transformed into the molecular form after subsequent cooling. Contrary to the classical view of the CMZ, recently, \citet{kru15a} and \citet{hen16a} showed that the observed position-velocity distribution of the dense gas in the CMZ follows an open stream rather than the $\mathrm{X}_2$-type, closed orbits. Many aspects of the CMZ are being explained based on this new point of view \citep{kru15a,kru15b,kru17,hen16a,hen16b}.

The accumulated molecular gas in the CMZ is estimated to be $2-5\times
10^7\msun$, comprising $5-10\%$ of the total molecular gas in the Galaxy
\citep{lau02,pie00,mor96,dah98,mol11}. The high densities of molecular gas in the CMZ lead to high star formation rates (SFRs) of $\sim0.1 \msun$/yr with active star formation sites (Sgr B, and Sgr C) and extraordinary, young, and massive star clusters (Arches and Quintuplet) \citep[among others]{yus08,yus09,fig99,kim06,imm12,lon13}. However, the SFR in the CMZ is an order of magnitude lower than predicted rates at high densities \citep{lon13}. Recent studies suggested that the low SFR in the CMZ is caused by the gas clouds in the CMZ undergoing episodic, bursty star formation, where the current stage is close to a minimum \citep{kru14,kru15b,kru17}.

Early numerical studies on gas dynamics in the Galactic 
bulge include two-dimensional (2D) sticky particle simulations by \citet{jen94}
and 2D smoothed particle hydrodynamics (SPH) simulations by \citet{lee99}
and \citet{eng99}.  They confirmed that the transition of gas motion from
X$_1$ to X$_2$ orbits indeed takes place in a bar potential, which was
originally suggested by \citet{bin91}. \citet{rod08} performed 2D sticky simulations
using density distribution models of the Galactic bulge, based on the 2MASS star
count map.  \citet[hereafter Paper I]{kim11} performed three-dimensional
(3D) SPH simulations of the formation of the CMZ in a simple $m=2$ bar with
a power-law density profile considering various astrophysical processes
such as heating/cooling of the gas, star formation, and supernova (SN)
feedback. More recently, \citet{kru15a} calculated orbital motions of gas clouds in the CMZ under the observed Galactic central mass profile of \citet{lau02}.
\citet{kru15b} showed that acoustic instabilities inside the Galactic inner Lindblad resonance (ILR; $\sim1$~kpc) account for the gas migration to the interior of the CMZ, $\sim100$~pc. \citet{kru17} extended this model to include effects of star formation, feedback, and stellar winds and showed that episodic, bursty star formations well reproduce the relatively low SFR in the CMZ compared to that expected in high densities.
These numerical studies have widened the knowledge on the formation
mechanism and other properties of the CMZ.  But these works implemented
simplified galactic potentials, narrow radial range of galactic potential and/or did not consider various astrophysical
processes in the gas mentioned earlier, and thus were not able to realistically
follow the exact gas motions from the Galactic disk to the CMZ region.

Ferrers ellipsoids \citep{fer87} have been commonly adopted in numerical
simulations as a density distribution model for the galactic bar component.
While Ferrers ellipsoids have, however, a rather unrealistic density cutoff, 
the exact infalling motion of gas from the galactic disk to the nucleus may 
be sensitive on the detailed galactic potential in the disk-bulge or disk-bar 
transition area.

In this paper, we perform 3D SPH simulations that consider
gas heating and cooling, star formation, and SN feedback
in order to trace the formation and evolution of the CMZ
under a realistic density distribution of the Milky Way. 
Instead of building up the Galaxy with simplified density
distribution models individually for the halo, disk, bulge, and bar,
we adopt a snapshot from a high-resolution simulation
targeted for the Milky Way, which was performed
by \citet[hereafter B15]{bab15}\footnote{The simulation of B15 reproduce the 
3D structures of the stellar disk, grand-design spiral arms, bar, and also 
atomic/molecular gas layers of the Milky Way using 6.4 million stellar and 4.5 
million gas particles. The dark matter halo component is set to be a static 
potential following a Navarro-Frank-White (NFW) profile \citep{nav97}}.
In order to simulate the infalling motion of gas from the Galactic disk to
the nucleus with a high spatial resolution and affordable computational overhead,
we develop a procedure based on the multipole expansion (ME) technique
to describe the detailed density distribution in the snapshot of the B15 simulation.
The ME method is advantageous for solving the Poisson equation
for a given density distribution from a simulation snapshot,
in that one can easily obtain density distributions
with bar/bulge masses and elongations that are slightly different from
the original density distribution by modifying
the expansion coefficients of the ME models. 

Since our simulation model is not designed to reproduce detailed structures and realistic kinematics of the dense gas clouds in the CMZ, in this study, we analyze our simulation results focusing on how far the gas clouds in the Galactic disk can reach into the central region under the realistic Galactic structure and how an asymmetric Galactic mass distribution affects the nuclear ring, rather than confining the CMZ to the X$_2$-type nuclear ring or an open stream (e.g., \citealt{kru15a,kru15b,kru17}).
This paper is organized as follows.
\S 2 explains how we describe the realistic Galactic structure
using the ME method. \S 3 presents our galaxy models and
parameters, and \S 4 describes the simulation code.
Our simulation results on the formation and evolution
of the CMZ are analyzed in \S 5.
Summaries and discussions are presented in \S 6.


\section{{Realistic Density Distribution Using Multipole Expansion Method}}

\subsection{Test of various multipole expansion models}

In the ME method,
the density $\rho$ and potential
$\Phi$ at a position $\vec{r}$ are expressed in terms of a series:
\begin{eqnarray}
	\rho(\vec{r}) = \sum_{nlm}A_{nlm}\rho_{nlm}(\vec{r}),\nonumber\\
	\Phi(\vec{r}) = \sum_{nlm}A_{nlm}\Phi_{nlm}(\vec{r}),
\end{eqnarray}
where $n$, $l$, and $m$ are quantum numbers for the radial and two angular variables, 
respectively, $A_{nlm}$ is the expansion coefficient, and $\rho_{nlm}$ and $\Phi_{nlm}$ 
are the density--potential basis sets of the ME method.

\begin{figure}
\centering
\includegraphics{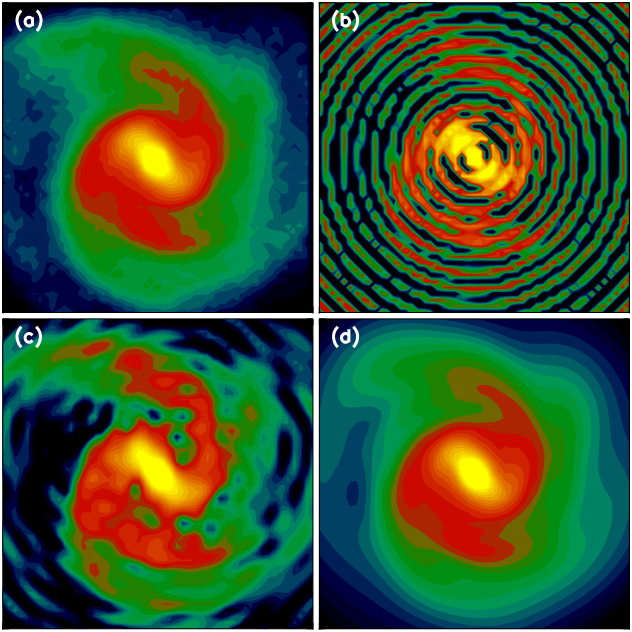}
\caption{Projected density distributions described with 6.4 million stellar
particles (a) and with various multipole expansion models (b-d). Each map
covers a $\pm10~$kpc rectangular region from the Galactic center.
(a) The snapshot from the B15 simulation at $T = 2.5$~Gyr. The Galactic
bar has grown substantially. (b) The E96 model, adopted for entire structures,
shows a distinct radial fluctuation. (c) The Hybrid model, HO92 for the 
nucleus+bulge and E96 for the disk component, better describes the 
Galactic structure, but still suffers a radial fluctuation. (d) The AI78 model in 
replacement of E96 model for the disk component best reproduces the overall 
density distribution of the adopted snapshot.}
\label{contour}
\end{figure}

To find the best set (or sets) of basis functions for the density distribution of the adopted
simulation snapshot (Figure~\ref{contour}a), we try the following three basis functions:
1) a 3D cylindrical basis set developed by \citet[hereafter E96]{ear96},
which is targeted for disk-like mass distributions with a non-zero
thickness, 2) a 3D spherical basis set developed by \citet[hereafter HO92]{her92},
which is optimized for quasi-spherical mass distributions, and 3) a 2D polar
basis set developed by \citet[hereafter AI78]{aok78} for a razor-thin disk.
These basis sets are listed in Table~\ref{tbl-1}.

First, we try the 3D cylindrical basis set of E96
to describe the density distribution of the B15 snapshot\footnote{Hereafter, the B15 
snapshot (or model) means the snapshot (or model) at $T=2.5$~Gyr if not specified otherwise.}.
Although we attempted many different combinations of maximum 
values for the three quantum numbers of $n$, $l$, and $m$, the density distribution 
that best fits the simulation snapshot has radial fluctuations as seen in 
Figure~\ref{contour}b. The problem appears to stem from the fact
that a basis function that is intended for a disk-like structure
is used to model a mass distribution having both disk and bulge components. 

\begin{figure}
\centering
\includegraphics{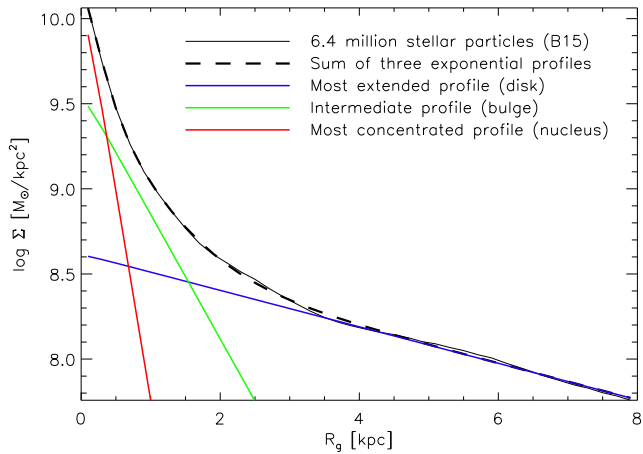}
\caption{Adopted surface density profile ($\Sigma$) for the 6.4 million stellar
particles of B15 (black solid line). The profile can be described with the sum of 
the three exponential profiles (black dashed line): The most extended profile 
(blue line) is regarded as the disk component, while the sum of the inner two exponential 
profiles (red and green lines) is regarded as the nucleus+bulge components.}\label{decomp}
\end{figure}

In an attempt to solve this problem, we decompose the mass distribution
of the simulation snapshot.
We find that
the projected density profile
of the snapshot is well decomposed into three parts using a double-exponential
profile (see Figure~\ref{decomp}),
\begin{equation}
	\Sigma(R,z) = \Sigma_0\exp(-R/R_d)[\exp(-|z|/h_z)/(2h_z)],
\end{equation}
where $\Sigma_0$, $R_d$, and $h_z$ are the central surface density, scale
length, and scale height, respectively.  Using the three double-exponential
profiles whose combined superpositions best fit the projected density profile of the
simulation snapshot, each particle in the simulation is assigned to one of
the three groups, disk (most extended), bulge (intermediate),
and nucleus (most concentrated), in a probabilistic manner (Table~\ref{tbl-2}).
Note that all of the resulting three groups have non-axisymmetric structures
(bar and spiral arms) although the fitting functions are axisymmetric.

\begin{table}
\caption{Decomposition of the galaxy model\label{tbl-2}}
\centering
\begin{tabular}{lccc}
\tableline
Component & $\Sigma_0$ & $R_d$ & $h_z$ \\
          & $[\msun/{\mathrm{kpc}}^2]$ &$[\mathrm{kpc}]$ & $[\mathrm{kpc}]$ \\
\tableline\tableline
Disk & $4.14\times10^8$ & 4.05 & 0.39 \\
Bulge & $3.88\times10^9$ & 0.59 & 0.42 \\
Nucleus & $1.68\times10^{10}$ & 0.18 & 0.16 \\
\tableline
\end{tabular}
\tablecomments{Three double-exponential profiles with these parameters are used to 
decompose the mass distribution of the simulation snapshot.}
\end{table}

We then apply the E96 basis set to the mass distribution made by the particles in the 
disk group and the 3D spherical basis set of HO92 to the collection of particles in the 
respective bulge and nucleus groups.
Figure~\ref{contour}c presents the projected
density distribution reproduced by the ME model
with a combination of the E96 and HO92 basis sets.
It shows less severe radial fluctuations than those seen in the
model with only the E96 basis set.
However, the projected density distribution
reproduced by the ME model still has unacceptably large discrepancies
from that of the simulation snapshot.
As in the case of the E96 basis set, we have tried many
different combinations of maximum values for $n$, $l$, and $m$ and
the scale factor for the radial basis function.  We find that larger
radial scale factors reduce radial fluctuations but describe
spiral structures in the disk less effectively.

Finally, we apply the 2D cylindrical basis set of AI78 to the mass distribution
of the disk group particles and the HO92 basis set to the distribution of the 
nucleus+bulge group particles.
Figure~\ref{contour}d displays the projected density distribution
reproduced by the ME model with a combination of HO92 and AI78 basis sets.
It shows an acceptably good match to the projected density
distribution of the B15 snapshot.
For this reason, we adopt these two basis sets
for describing the mass distribution of the simulation snapshot
with the ME technique.
The two-dimensional limit of the AI78 basis set
is alleviated by the implementation of the ``thick disk approximation'',
which will be described in \S~2.3.

\subsection{Expansion coefficients}

We calculate two sets of expansion coefficients for the particle distribution of the 
B15 snapshot: one with the HO92 model and the other with the AI78 model.
In Appendices~A and B, we show how the HO92 and AI78 models describe the 
density and potential distributions, how the expansion coefficients can be obtained from 
a given particle distribution, and how the gravitational acceleration $\vec{a}$ can be 
obtained from these ME models.

Hereafter, $n$ and $m$ corresponds to radial and azimuthal quantum numbers in 
both the HO92 and AI78 models, while $l$ is polar quantum number in the HO92.
Higher quantum numbers represent smaller structures and higher frequencies. However, 
including higher quantum numbers does not always result in a better description of a given 
density distribution, because they may yield noisier structures (e.g., \citealt{mei14}).  
Furthermore, the use of higher quantum numbers requires significantly longer calculation 
times to derive the acceleration $\vec{a}$.  Thus, one needs to find the appropriate maximum 
quantum numbers $n_{max}$, $l_{max}$, and $m_{max}$ that can properly describe the fine 
structures of the given particle distribution and yet do not results in an excessive calculation cost. 
Among the possible combinations of $n_{max}$, $l_{max}$, and $m_{max}$, we selected the one that satisfied the following criteria: (1) the projected density maps derived with the stellar particles and the ME models (cell side length = 0.2~kpc) have a relative difference smaller than 0.1 in the radial range of $R_g<6$~kpc, and (2) $n_{max}$, $l_{max}$, and $m_{max}$ have the smallest number of coefficients $A_{nlm}$. We tested a few thousand sets of $n_{max}$, $l_{max}$, and $m_{max}$ and found that $n_{max}=22$, $l_{max}=10$, and $m_{max}=10$ ($n_{max}=22$ and $m_{max}=10$
for the AI87 model) were the best choice for describing the nucleus, the bar-like bulge and the spiral-structured disk, with a total number of coefficients $A_{nlm}$ of 2783 for the HO92 model and 253 for the AI78 model. 

\begin{figure}
\centering
\includegraphics{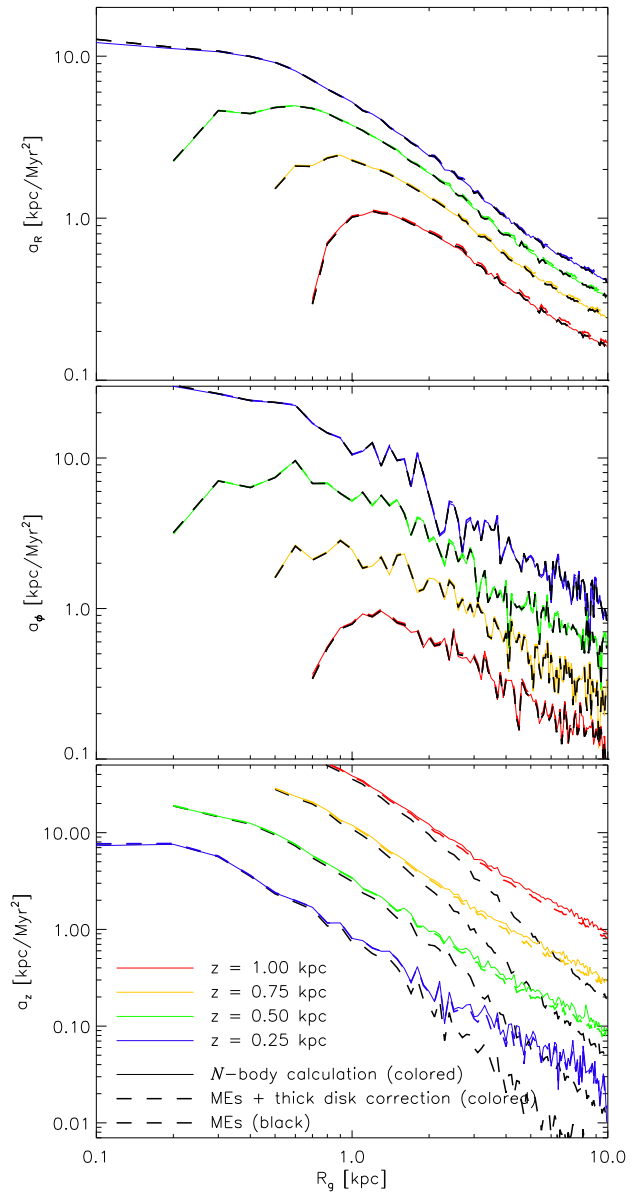}
\caption{$a_R$, $a_{\phi}$, and $a_z$ profiles of the Galaxy model. Colored
solid lines are from an $N$-body calculation of B15, while colored dashed lines are
derived with the two multipole expansion models (HO92 + AI78)
and thick disk correction. Black dashed lines represent the acceleration
profiles before the thick disk correction. For illustration purposes,
we arbitrarily shift the profiles along the y-axis.}\label{accel}
\end{figure}

\subsection{Thick disk approximation}

Since we describe the disk component with the 2D polar model of AI78,
effects of the vertical mass distribution on the acceleration, $\vec{a}$, have 
thus far been ignored. The disk component is decomposed to follow the 
exponential disk profile of Equation~(2), and thus we can estimate the effect of 
the vertical structure of the disk on the radial and altitude accelerations, $a_R$ and
$a_z$, following the ``thick disk approximation'' \citep[see Section 2.6 for details]{bin08}. 
First, we calculate the idealized potentials of razor-thin (2D) and non-zero thickness (3D) 
exponential disks as
{\small\begin{eqnarray}
	\Phi_0(R) &=& -\pi G \Sigma_0 R\left[I_0(y)K_1(y)-I_1(y)K_0(y)\right],
		       \nonumber\\
	\Phi(R,z) &=& -\frac{4G\Sigma_0}{R_d}\int_{-\infty}^{\infty}dz'\frac{
		       \exp(z'/h_z)}{2h_z}\times\nonumber\\
		  & &  \int_{0}^{\infty}da~\sin^{-1}\left(\frac{2a}{\sqrt{+}
		       +\sqrt{-}}\right)aK_0(a/R_d),
\end{eqnarray}}
where $I_n$ and $K_n$ are modified Bessel functions, and $y$ and $\sqrt{\pm}$
are defined as
\begin{eqnarray}
	y          &\equiv& R/(2R_d),\nonumber\\
	\sqrt{\pm} &\equiv& \sqrt{(z-z')^2+(a\pm R)^2}.
\end{eqnarray}
Then the difference of $a_R$ between the 2D and 3D exponential disks is
derived as
\begin{eqnarray}
	\triangle{a_R(R,z)} &=&  a_R(R,z)-a_{R,0}(R)\nonumber\\
			    &=& -d\Phi(R,z)/dR+d\Phi_0(R)/dR.
\end{eqnarray}
By adding $\triangle{a_R(R,z)}$ to the $a_R$ derived from the AI78 model
(Equation~(B5)), we can recover the effect of the non-zero disk thickness on $a_R$.
Unlike the $a_R$ calculated with the AI78 model, $\triangle{a_R(R,z)}$ is
not a function of the azimuthal angle $\phi$, since the idealized exponential disk
profile is axisymmetric. Thus, $\triangle{a_R(R,z)}$ is equally applied
for $\phi=0-2\pi$, and we assume that the effect of non-zero disk thickness
is negligible on $a_{\phi}$. Meanwhile, $a_z$ due to the non-zero disk
thickness can be assigned to be
\begin{equation}
	a_z(R,z) = -d\Phi(R,z)/dz,
\end{equation}
for $\phi=0-2\pi$.

\subsection{Implementation}

Accelerations derived by the ME method are compared with the $N$-body
calculation in Figure~\ref{accel}. To reduce the local fluctuations, $a_R$, $a_{\phi}$,
and $a_z$ are averaged for a given cylindrical shell. The averaged
acceleration profiles that are constructed with the ME models with and
without the thick disk correction effectively reproduce that of an $N$-body
calculation for a wide range of $R_g$ and $z$.  The thick disk correction plays 
an important role in reproducing $a_z$ for $R_g \gsim 2~$kpc.  Small
discrepancies are observed in $a_z$ at $R_g\gsim3~$kpc.
However, these would influence our scientific results only negligibly,
since we focus on the gas motions along the Galactic plane for
$R_g\lsim3~$kpc.

The realistic Galactic structure described with the ME method and the
thick disk correction are included in our SPH simulations as a fixed
Galactic potential. (The simulation code is described in \S~4.)
Since our galaxy model contains non-axisymmetric
structures such as the elongated bar and spiral arms, we need to consider
their rotation. The pattern speed of the bar ($\Omega_{bar}$)
of the B15 model is estimated to be $\sim35~$km/s/kpc,
and the co-rotation radius ($R_{\mathrm{CR}}$) is inferred to be $\sim5.5~$kpc.
We assume that the spiral arms in all
$R_g$ ranges rotate with a constant angular velocity that is the same as
$\Omega_{bar}$. Thus, the whole Galactic structure is considered to rotate
with the same pattern speed. A spherically symmetric
potential for the halo component is added on top of the rotating
non-axisymmetric structures.
For this, we adopt the same NFW profile
used in B15, which has a total mass of $1.26\times10^{12}\msun$, a scale radius
of 280~kpc, and a concentration of 11.2.

\begin{figure*}
\centering
\includegraphics{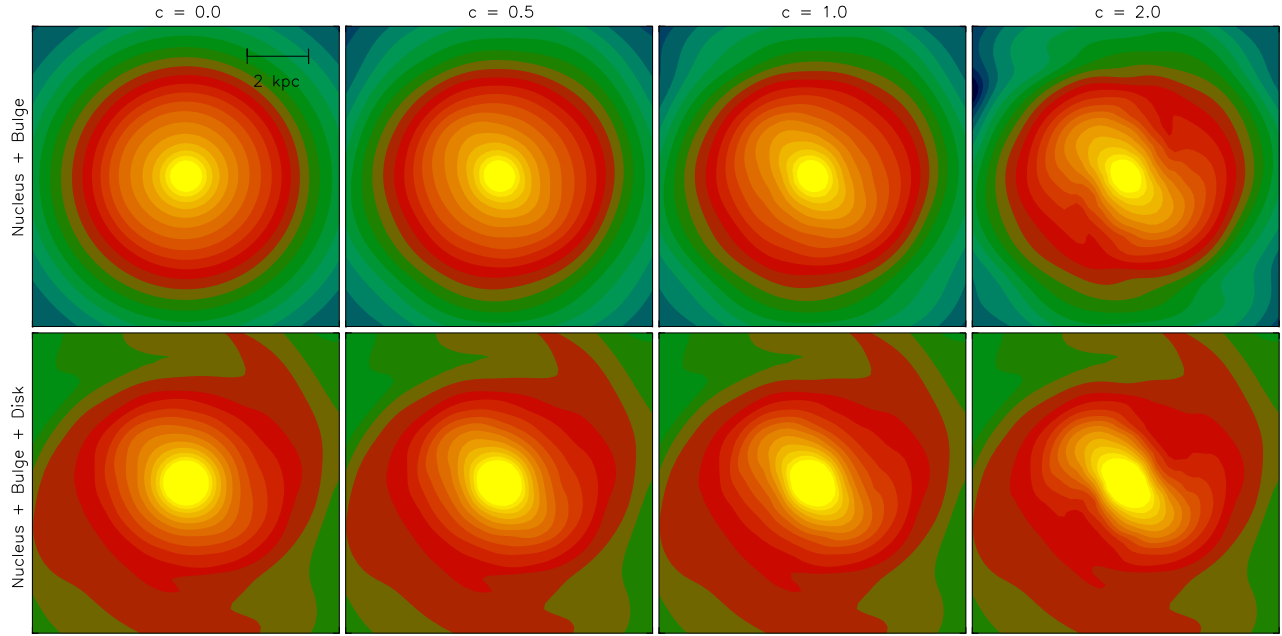}
\caption{Projected density distributions constructed with the modified
coefficients set of $c=0.0$, 0.5, 1.0, and 2.0 for $\pm~5$~kpc rectangular region from the Galactic center. For a clear appearance,
different color codings are used for the nucleus+bulge (upper panels) and
nucleus+bulge+disk (bottom panels) components.}\label{strength}
\end{figure*}

\section{Galaxy Models}

\subsection{Bar structure}

As mentioned earlier, one of the advantages of the ME method is that
more and less elongated structures can be easily generated by adjusting 
the expansion coefficients. In order to vary the bar elongation in our galaxy model,
we multiply $m = 2, 4$ coefficients of the nucleus+bulge component
by a constant $c$. All coefficients for the nucleus+bulge component
are then rescaled in a way that reproduces the original total mass.
Figure~\ref{strength} shows the projected density distribution of the
nucleus+bulge and nucleus+bulge+disk components,
constructed with different $c$ values of 0.0, 0.5, 1.0, and 2.0.
It is apparent from the figure that the nucleus+bulge component becomes more 
elongated as the value of $c$ increases. At $c=2.0$, the nucleus+bulge is not only 
more elongated than at $c=1.0$, but also exhibits spiral patterns.
Since the spiral patterns weakly remain in the bulge 
component even after the decomposition,
the nucleus+bulge with the amplified $m = 2, 4$ coefficients
reveals the spiral patterns.
The more/less elongated nucleus+bulges are smoothly connected with
the spiral patterns without abrupt density changes (see the bottom panels of Fig~\ref{strength}).
However, the bar elongation cannot be completely removed
even when $c=0.0$,
since the disk component includes some of the elongated structure as well.

The bar strength of our models is measured by a dimensionless parameter $Q$, 
which is defined as 
\begin{equation}
	Q \equiv {\frac{a_{\phi}(R,\phi)}{a_R(R)}}\mid_{max},
\end{equation}
\citep{com81}.
For computational convenience, $a_R$ and $a_{\phi}$ include the nucleus and bulge components only. The galaxy model from B15 (at $T = 2.5$~Gyr) is estimated to have a bar strength of $Q = 0.131$.

In some of our model runs, the bar is set to grow gradually corresponding to 
the bar growth timescale $\tau_{growth}$ from the start of the simulations.
We imitate the gradual bar growth using a series of coefficient sets that are 
generated with 20 equally spaced $c$ values from 0 to $c_{max}$. For a given 
bar growth timescale, $c$ is linearly interpolated with the simulation time-step,
and a different coefficient set is adopted accordingly. Bar strength is fully turned on 
by the end of $\tau_{growth}$, and thus the coefficient set of $c_{max}$ is used for 
all times after $\tau_{growth}$.

\begin{deluxetable}{lcccccc}
\tabletypesize{\footnotesize}
\tablecaption{Model parameters\label{tbl-3}}
\tablehead{
\colhead{Model} & \colhead{$N_{gas}$} &
\colhead{$R_{\mathrm{IB}}-R_{\mathrm{OB}}$} & \colhead{Q} & \colhead{$c_{max}$} &
\colhead{$\tau_{growth}$} & \colhead{$\Omega_{bar}$}  \\
\colhead{} & \colhead{} &
\colhead{[kpc]} & \colhead{} & \colhead{} &
\colhead{[Myr]} & \colhead{[km/s/kpc]}}
\startdata
A    & $10^5$ 		& 0 - 6 & 0.131 & 1.00&    0 & 35   \\
Aq1  & $10^5$	 	& 0 - 6 & 0.081 & 0.50&350 & 35   \\
Aq2  & $10^5$	 	& 0 - 6 & 0.106 & 0.75&350 & 35   \\
Aq3  & $10^5$	 	& 0 - 6 & 0.157 & 1.25&350 & 35   \\
Aq4  & $10^5$	 	& 0 - 6 & 0.184 & 1.50&350 & 35   \\
At1  & $10^5$	 	& 0 - 6 & 0.131 & 1.00&175 & 35   \\
At2  & $10^5$	 	& 0 - 6 & 0.131 & 1.00&350 & 35   \\
At3  & $10^5$	 	& 0 - 6 & 0.131 & 1.00&700 & 35   \\
Ap1  & $10^5$	 	& 0 - 6 & 0.131 & 1.00&350 & 30   \\
Ap2  & $10^5$	 	& 0 - 6 & 0.131 & 1.00&350 & 40   \\
B    & $10^5$	 	& 1 - 6 & 0.131 & 1.00&    0 & 35   \\
Bq1  & $10^5$	 	& 1 - 6 & 0.081 & 0.50&350 & 35   \\
Bq2  & $10^5$	 	& 1 - 6 & 0.106 & 0.75&350 & 35   \\
Bq3  & $10^5$	 	& 1 - 6 & 0.157 & 1.25&350 & 35   \\
Bq4  & $10^5$	 	& 1 - 6 & 0.184 & 1.50&350 & 35   \\
Bt1  & $10^5$	 	& 1 - 6 & 0.131 & 1.00&175 & 35   \\
Bt2  & $10^5$	 	& 1 - 6 & 0.131 & 1.00&350 & 35   \\
Bt3  & $10^5$	 	& 1 - 6 & 0.131 & 1.00&700 & 35   \\
Bp1  & $10^5$	 	& 1 - 6 & 0.131 & 1.00&350 & 30   \\
Bp2  & $10^5$	 	& 1 - 6 & 0.131 & 1.00&350 & 40   \\
Ah   & $3\times10^5$	& 0 - 6 & 0.131 & 1.00&    0 & 35   \\
At2h & $3\times10^5$	& 0 - 6 & 0.081 & 0.50&350 & 35   \\
Bh   & $3\times10^5$	& 1 - 6 & 0.131 & 1.00&    0 & 35   \\
Bt2h & $3\times10^5$	& 1 - 6 & 0.081 & 0.50&350 & 35   \\
\enddata
\tablecomments{The standard values of $Q$, $c_{max}$, $\tau_{growth}$, and
$\Omega_{bar}$ are 0.131, 1.0, 350~Myr, and 35~km/s/kpc,
respectively (see text for details). }
\end{deluxetable}

\subsection{Model differences}

We construct a number of galaxy models, as listed in Table~\ref{tbl-3}. The models 
are divided into two types, A and B. In models of type A, gas particles are distributed 
on the disk all the way down to $R_g=0~$kpc, the inner boundary of the initial gas 
distribution, $R_{\mathrm{IB}} = 0$. In models of type B, the gas particles are distributed 
down to $R_{\mathrm{IB}} = 1$~kpc. The reason we test the two different values of 
$R_{\mathrm{IB}}$ is to examine how the initial existence or nonexistence of the inner 
gas structure affects the evolution of the CMZ. In both types, the outer boundary of the 
initial gas distribution $R_{\mathrm{OB}}$ is set to 6~kpc. This is slightly larger than 
$R_{\mathrm{CR}}$, the boundary between inward and outward radial motion of particles.

Both types of our models are further divided depending on the different values of
bar strength $Q$ (Aq1(Bq1) through Aq4(Bq4)), the bar growth time scale $\tau_{growth}$ 
(At1(Bt1) through At3(Bt3)), and the pattern speed of the bar $\Omega_{bar}$ 
(Ap1(Bp1) and Ap2(Bp2)). In addition, four of our models are built with a higher resolution 
(Ah, At2h, Bh, and Bt2h). In said high-resolution models, the number of gas particles 
(whose individual mass $m_{gas}$ is $2.5-2.8\times10^3\msun$) distributed on the disk is 
$N_{gas} = 3\times10^5$. In the other models (except Ah, At2h, Bh, and Bt2h), the number 
of gas particles, whose $m_{gas}$ is $7.5-8.4\times10^3\msun$, is $N_{gas} = 10^5$. For a 
given particle number, $m_{gas}$ is adjusted to reproduce the total gas mass of the B15 model, 
over the range of $R_{\mathrm{IB}}\le R_g\le~R_{\mathrm{OB}}$. Here, $m_{gas}$ of the 
high-resolution models is slightly smaller than that of the B15 model, $3\times10^3\msun$.
A comparison between our high and low-resolution runs shows a good agreement at 
$T=2~$Gyr within $\sim10\%$.

In models A, Ah, B, and Bh, the bar strength and the pattern speed of the bar are respectively 
set to $Q = 0.131$ and $\Omega_{bar} = 35$~km/s/kpc, which are equal to those of the B15 
model (``standard" values). The bar strength in these runs is fully turned on at the beginning 
when $\tau_{growth} = 0$ (standard value), and the initial positions ($\vec{r}$) and velocities 
($\vec{v}$) of the gas particles are adopted from the simulation snapshot of B15. Thus the initial 
gas distribution reflects the realistic gaseous structures (spiral arms, dust lanes, and nuclear ring) 
that are hydro-dynamically evolving under the Galactic structure.

Unlike in runs A, Ah, B, and Bh, $Q$ is gradually turned on for $\tau_{growth}$
in all the other runs. Three different $\tau_{growth}$ timescales of 175, 350, and 700~Myr
are chosen in runs At1(Bt1), At2(Bt2), and At3(Bt3), respectively. These values correspond 
to 1, 2, and 4 times the rotation period of $\tau_{bar}\sim175~$Myr for the standard 
$\Omega_{bar}$ of 35~km/s/kpc. Using four different values of $c_{max}$, 0.50, 0.75, 
1.25, and 1.5, we generate the more/less elongated bar structure whose full $Q$ is 0.081, 
0.106, 0.157, and 0.184, respectively. These $Q$ are set in runs of Aq1(Bq1) through Aq4(Bq4), 
respectively, and are gradually turned on for $\tau_{growth}$ = 350~Myr. Two different 
$\Omega_{bar}$ values of 30 and 40~km/s/kpc are tested in runs Ap1(Bq1) and Ap2(Bq2), 
respectively, while the standard $\Omega_{bar}$ of 35~km/s/kpc is adopted in the others. 
Since the runs with $\tau_{growth} > 0$ start with the least elongated bar structure with $c=0.0$,
the initial conditions from the B15 model are no longer realistic. For these runs, the initial 
$\vec{r}$ of each gas particle is generated to follow a flat radial profile and a vertical distribution 
of the Gaussian function with a scale height of 20~pc, which approximates the B15 gas disk. 
The initial $\vec{v}$ is assigned to follow a circular orbit, while the vertical component is initially 
set to be zero. 

\section{Simulation code}

For our hydrodynamic simulations, we use a parallel $N$-body/SPH code,
Gadget-2 \citep{spr05}. The softening length is fixed to be 30~pc, and the
size of the SPH kernel\footnote{The size of the SPH kernel is limited to describing gas density only up to $\sim200\msun/\mathrm{pc}^3$.} is variable by imposing the smoothed number of neighbors
to be $32\pm2$. For a higher performance with each time-step, we adopt the 
``FAST'' algorithm of \citet{sai10}, which assigns separate time-steps for gravitational
and hydrodynamical integration. A realistic Galactic structure is considered 
an external potential using the ME method as described in \S 2, and thus only
the gas disk is initially composed of simulation particles
(as described in \S~2 and 3).

We implement realistic astrophysical gas processes of 1) radiative
heating/cooling, 2) star formation, and 3) stellar feedbacks \citep{shi14}.
Using the CLOUDY 90 package \citep[version 10.10]{fer98}, we
calculate the radiative cooling and heating rates for wide density and temperature
ranges of $10^{-6}-10^{6}~\mathrm{H/cm^3}$ and $10-10^8~$K under a uniform far-ultraviolet radiation (FUV) field
of 1 and $100~G_0$, where $1~G_0$ is the observed FUV
strength in the solar neighborhood\footnote{$G_0\equiv1.6\times10^{-3}~\mathrm{erg~cm^{-2}~s
^{-1}}$ is the Habing field \citep{hab68}}. The FUV strength is a function of atomic
hydrogen column density, stellar number density, and stellar age, and thus it
is time-varying and also position-dependent. In this study, we assume that
intensive star formation activities in the $R_g\leq1~$kpc region produce 
constant FUV radiation of $100~G_0$, while those in the outer region produce constant 
FUV radiation of $1~G_0$.

We transform a gas particle into a star particle when it satisfies all of
the following star formation criteria \citep{sai08,sai09b}: 1) a hydrogen number 
density of $n_H>100~\mathrm{cm}^{-3}$, 2) temperature
of $T<100$~K, and 3) convergent flow of $\nabla\cdot v<0$. The local star
formation rate is calculated according to the Schmidt law,
\begin{equation}
	\frac{d\rho_*}{dt} = c_{*}\frac{\rho_{gas}}{t_{dyn}},
\end{equation}
where $\rho_{*}$ is the density of newly formed stars for a given time step
$dt$, $c_*$ is the characteristic star formation efficiency of 0.033,
$\rho_{gas}$ is the local gas density, and $t_{dyn}$ is the local dynamical
time.  Here, each stellar particle represents a star cluster whose stellar
population follows the Kroupa mass function in the mass range of $0.1-100\msun$
\citep{kro01}.

Using the stellar evolutionary tracks of \citet{hur00}, we
calculate the number of massive stars that may eventually end up as
SN$_{\mathrm{II}}$ explosions. SN$_{\mathrm{II}}$ feedback is considered in
a probabilistic manner, following the recipe of \citet{oka08}.  Energy, mass, and
metal ejected by SN$_{\mathrm{II}}$ explosions are
distributed to neighboring gas particles. To better capture the SN shock
front with the individual time-stepping \citep{spr05}, we implement a time-step 
update algorithm introduced by \citet{dur12}, which is a modified version of a 
time-step limiter of \cite{sai09}. Stellar winds from less massive stars are treated to continuously eject momentum, mass, and metals into the surrounding neighboring gas particles. The details for the radiative heating/cooling, the star formation and stellar feedback are described in \citet{shi14} and references therein. 

\section{Results of the simulations}

\subsection{Performance}

We run all of our simulations for 2000~Myr. Each of the high-resolution runs takes 
$\sim14$ days for $T=2000~$Myr using 32 cores, while the self-consistent Galaxy 
simulation of B15 takes $\sim1$ month for $T=4500$~Myr using a thousand cores.
Calculation of our high-resolution runs is estimated to be $\sim30$ times faster
than that of B15, in which $m_{gas}$ is even $\sim10\%$ smaller.
The high performance arises mainly because the ME replaces the B15's 6.4 
million stellar particles and the gas distribution is truncated to a much smaller 
$R_{\mathrm{OB}}$ than that of the B15 model, thus reducing the total number 
of gas particles to $\sim66\%$ of B15. Thanks to the high performance of the ME 
method, we are able to perform the hydrodynamic simulations under the realistic 
Galactic structure with various model parameters.

\subsection{Overall morphology}

\begin{figure*}
  \centering
  \includegraphics{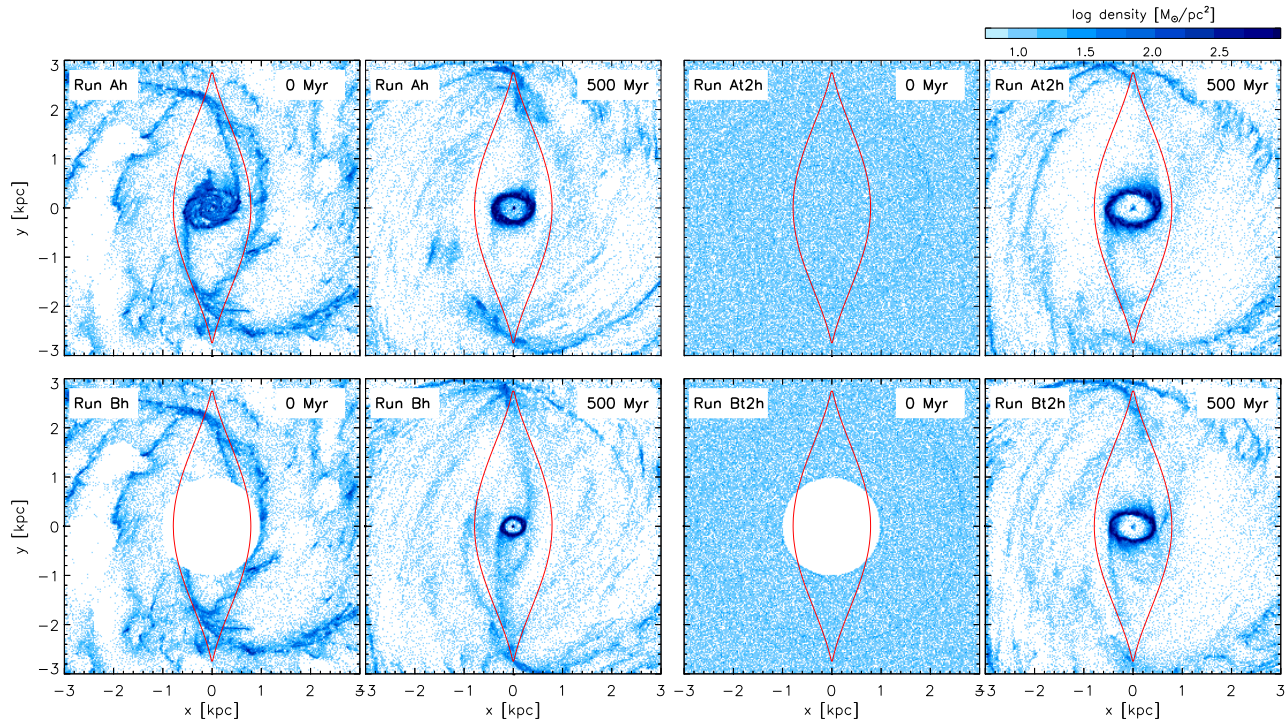}
  \caption{Projected density distributions at $T=0$ and 500~Myr for the inner 3~kpc
rectangular region from the four high-resolution runs (Ah, Bh, At2h, and Bt2h).
In each panel, the major axis of the bar structure is aligned along the y-axis.
The grey scale represents the gas density varying linearly from 5 to
$900\msun/\mathrm{pc}^{2}$. The innermost non-looped $\mathrm{X}_1$ orbit
is over-plotted with a solid curve (red). The two leftmost columns show the
gaseous structure at $T=0$ and 500~Myr from runs~Ah and Bh where the bar
structure is fully turned on at the beginning, while the two rightmost columns
show that the gaseous structure from runs~At2h and Bt2h where the bar structure 
grows gradually for $\tau_{growth}=350~$Myr.} \label{snapshot}
\end{figure*}

Figure~\ref{snapshot} presents the projected density distributions
of the gas clouds for the inner 3~kpc rectangular region in the high-resolution 
runs at $T=0$ and 500~Myr. Since the rotation of the non-axisymmetric Galactic 
structure is considered in the inertial frame, we redistribute the gas particles in a 
way that the bar major axis points toward the y-axis. The red curve over-plotted in 
each panel of the figure indicates the innermost $\mathrm{X}_1$ orbit that does not 
self-intersect. Its long axis is aligned with the bar's major axis, where its cusps are extended 
to $R_g\sim2.8~$kpc. The gas distribution at $T=0~$Myr in run Ah shows gaseous 
structures that have self-consistently evolved with the Galactic structures (B15). The gas 
streams along the spiral arms are smoothly connected with the dust lanes near 
$R_g\sim2.5~$kpc, i.e., the contact points with the bar major axis. Due to more significant 
angular momentum loss near the contact points and the dust lanes, the gas clouds plunge to the nuclear 
region and form a nuclear ring at $R_g\sim420~$pc. 

Unlike some claims that a nuclear ring forms near the ILR \citep{kna95,reg03,com10}, the radius of the nuclear ring at $\sim420$~pc is much smaller than the ILR, which is estimated to be $\sim1.2$~kpc in our Galactic model. According to hydrodynamic simulations performed by \citet{kim12}, the radius of the nuclear ring is determined by the amount of angular momentum loss at the dust lanes rather than by the resonance. Recently, \citet{kru15b} showed that gas clouds within the ILR can lose angular momentum due to acoustic instability and thus move closer to the galactic center, down to a stalling radius where shear is at a minimum.
Hereafter, a nuclear ring structure produced by our simulations is referred to as a ``nuclear ring", 
while the real, observed one is referred to as the ``CMZ''. 

The overall morphologies at $T=500~$Myr in run Ah are quite similar to the initial state, 
although the dust lanes extend farther and the nuclear ring becomes smaller than at 
$T=0~$Myr. The slight difference in the morphologies stems from the fact that the gas 
distribution used as the initial condition is still in the process of adjusting for the time-varying 
Galactic structure of B15. As the gas clouds are increasingly relaxed under the given Galactic 
structure, the starting points of the dust lanes become closer to the $\mathrm{X}_1$ cusps. 

The nuclear structure of $R_g<1~$kpc is initially removed in model~Bh. The inflowing gas 
clouds do not collide with the nuclear ring structure and thus approach more closely to the 
Galactic center, first forming an unstable $\mathrm{X}_3$-type orbit for a while. After the 
$\mathrm{X}_3$-type structure collapses, a stable $\mathrm{X}_2$ orbit instead settles 
down in the nuclear region. Despite the fact that the same Galactic potential is adopted
in both models~Ah and Bh, the existence/nonexistence of the nuclear ring at the beginning 
changes the nuclear ring size $R_{ring}$ at $T=500~$Myr.

Results from runs~At2h and Bt2h are shown together in Figure~\ref{snapshot} as well.
The overall morphology at $T=500~$Myr is similar to that of runs~Ah and Bh
except that the nuclear ring is larger. Since the full $Q$ is gradually turned on for 
$\tau_{growth}=350~$Myr in runs~At2h and Bt2h, the gas clouds infalling before 
$T=350~$Myr have a larger angular momentum than that of runs~Ah and Bh, 
and thus form the larger-sized nuclear ring. Although the same Galactic potential is 
adopted for runs~At2h and Bt2h, the nuclear ring that forms under run~At2h is slightly 
larger than that of run Bt2h.

\begin{figure}
  \centering
  \includegraphics{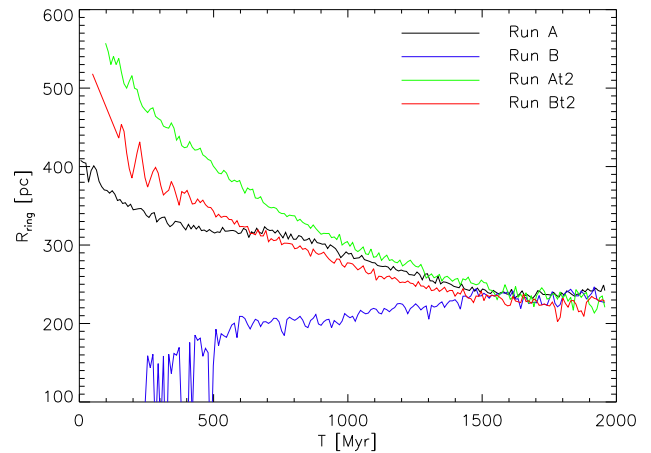}
  \caption{Temporal evolution of $R_{ring}$ from runs~A, B, At2, and Bt2.
  Here, $R_{ring}$ is calculated by a Gaussian peak of $R_g$ distribution for
  the gas clouds in the range $10<R_g<1000~$pc.  }
  \label{size_t}
\end{figure}

\begin{figure*}
\centering
\includegraphics{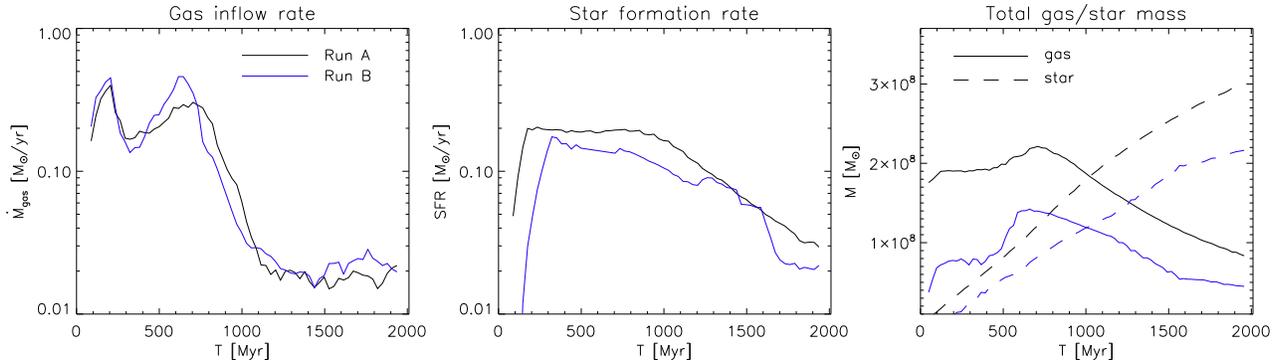}
\caption{Temporal evolution of the gas inflow rate to the nuclear ring
 $\dot{M}_{gas}$, the SFR in the nuclear ring, and the total masses of the
gas and the stars within the nuclear ring $M_{star}$ and $M_{gas}$ in runs~A and B.}\label{inflow}
\end{figure*}

\begin{figure}
\centering
\includegraphics{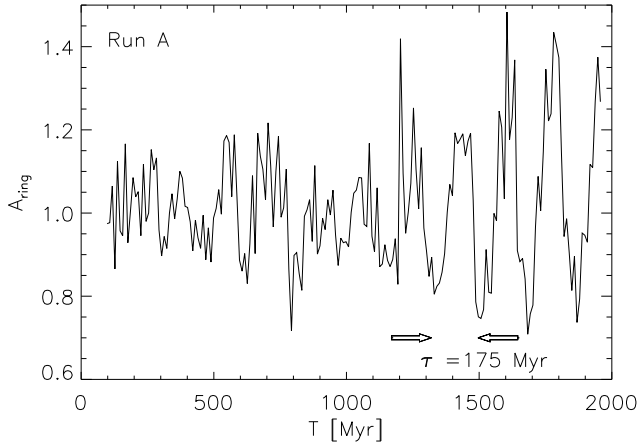}
\caption{Evolution of $A_{ring}$ in run~A. $A_{ring}$ oscillates with
a period of $\sim$175~Myr, which is the same as the Galaxy's
rotational period, $\tau_{bar}=175$~Myr.}
\label{asymmetry}
\end{figure}

\subsection{Temporal evolution}

Figure~\ref{size_t} shows the temporal evolution of the typical size of
the nuclear ring $R_{ring}$ for runs~A, B, At2, and Bt2.
Here, $R_{ring}$ is calculated by a Gaussian peak of $R_g$ distribution
for the gas clouds located in a range of $10<R_g<1000~$pc. Hereafter,
properties of a nuclear ring are measured in the same radial range.
Due to shear viscosity between the resident gas clouds in the nuclear ring and collisions between inflowing gas clouds and the resident gas clouds, the typical angular momentum and/or orbital energy of the nuclear ring change gradually, and $R_{ring}$ varies accordingly. All values of $R_{ring}$, regardless of the initial value, converge to the same value of $\sim240$~pc at $T\sim1500$~Myr, since the amounts of angular momentum of inflowing gas clouds along the dust lanes after $T\sim350$~Myr are the same. Interestingly, the equilibrated ring size at $\sim240$~pc is similar to the stalling radius \citep{kru15b}, which is calculated as $\sim270$~pc in our Galactic model. 

The inflowing gas clouds along the dust lanes are accumulated in the nuclear ring with a typical surface density of $\Sigma_{gas}\sim500-600\msun/\mathrm{pc}^2$. The gas clouds have a clumpy structure with an individual mass of $1-3\times10^6\msun$. The mass scales are comparable with those in the outer region of the CMZ ($R_g\gsim100~$pc) \citep{sta91}, but are $\sim$10 times larger than those in the inner region of the CMZ ($R_g\lsim100~$pc), whose surface density is $\Sigma_{gas}=1000-3000\msun/\mathrm{pc}^2$ \citep{wal15,hen16b}. The different mass scale of the gas clouds between the nuclear ring and the inner region of the CMZ might be due to the two facts that: 1) the mass scale for gravitational collapse against thermal pressure and shear force is different, since $R_g$ is different \citep{hen16b}, and 2) our simulation model does not realistically describe clumpy structures, since the minimum size of the SPH kernel is set to smooth the high-density clumpy structures of $\rho\gsim200\msun/\mathrm{pc}^3$.

The evolution of the gas inflow rate
to the nuclear ring ($\dot{M}_{gas}$), the SFR in the nuclear ring,
and the total masses of the gas and the stars within the nuclear ring
($M_{gas}$ and $M_{star}$) in runs~A and B are shown in Figure~\ref{inflow}.
During the first $\sim200~$Myr, unrelaxed gas clouds
under the given Galactic structure inflow to the nuclear ring (Paper I).
After the initial relaxation period, $\dot{M}_{gas}$ is enhanced again
as the gas clouds located in the outside $\mathrm{X}_1$ orbit inflow to
the inside through the spiral shocks \citep{kim14,seo14}.
After the second peak, the inflowing of gas clouds in the nuclear region
is rapidly decreased, and equilibrates to $\dot{M}_{gas}\sim0.02\msun/$yr
at $T\sim1200~$Myr in both runs. Without additional fresh gas supplements to the gas disk, such as gas recycling from stellar mass loss \citep{seg16}, galactic fountains \citep{opp08,opp10,ker09}, and cosmic accretion of primordial gas \citep{dek09,ric12}, the gas reservoir supplying the nuclear ring is depleted; thus, $\dot{M}_{gas}$ remains steady at $\dot{M}_{gas}\sim0.02\msun/$yr until the end of the runs.

Star formation activities in the nuclear ring show milder changes as compared to those of $\dot{M}_{gas}$ (Figure~\ref{inflow}). Since $\rho_*$ is proportional 
to $\rho_{gas}$ (Equation~(8)), SFR integrated for the entire nuclear ring is more correlated with $M_{gas}$ than $\dot{M}_{gas}$. 
As the accumulated gas clouds in the nuclear ring are consumed to form stars at an SFR of $\sim0.1\msun/\mathrm{yr}$, the gas densities in the nuclear ring decrease. Consequently, the SFR is decreased below $0.1\msun/\mathrm{yr}$.

\subsection{Asymmetry \& Tilt of the CMZ}

The Galaxy model used in this study reflects the realistic mass distribution that
self-consistently reproduces the 3D structure of the Galactic stellar disk, the
grand-design spiral arms, and the bar (B15). The realistic Galactic structures 
of B15 are found to have a lopsided central mass distribution, which leads the density peak 
of the central mass distribution to be off-centered from the Galactic center. Since the 
nuclear ring is centered on the off-centered density peak, not on the Galactic center, 
the center of the nuclear ring wanders around the Galactic center with the Galaxy's 
rotational period \citep{bal80,hay98,fux01,bou05}. The off-centered nuclear ring with 
respect to the Galactic center brings an asymmetric mass distribution $A_{ring}$, 
which is defined as the ratio of $M_{gas}$ with $x>0$ to that with $x<0$, where 
the $y$-axis is directed toward the Galactic center along the bar's major axis, 
the $x$-axis is along the Galactic plane, and the $z$-axis is perpendicular to the 
Galactic plane. As the center of the nuclear ring wanders around the Galactic center, 
a projected nuclear ring onto the $x$-$z$ plane goes back and forth along the $x$-axis.
Thus, $A_{ring}$ oscillates accordingly with a period of $\sim175~$Myr, which is the
same as the period of Galactic rotation, $\tau_{bar}=$175~Myr (Figure~\ref{asymmetry}). 
The amplitude of $A_{ring}$ gradually grows as $R_{ring}$ decreases to
$\sim240~$pc. 
However, the observed asymmetry of the CMZ is reported to have originated from a discontinuous, lopsided gas mass distribution whose one side of $l>0$ contains much more gas than the other side of $l<0$, rather than from the off-centered nuclear ring \citep{bal10,hen16a}.

\begin{figure*}[!t]
\centering
\includegraphics{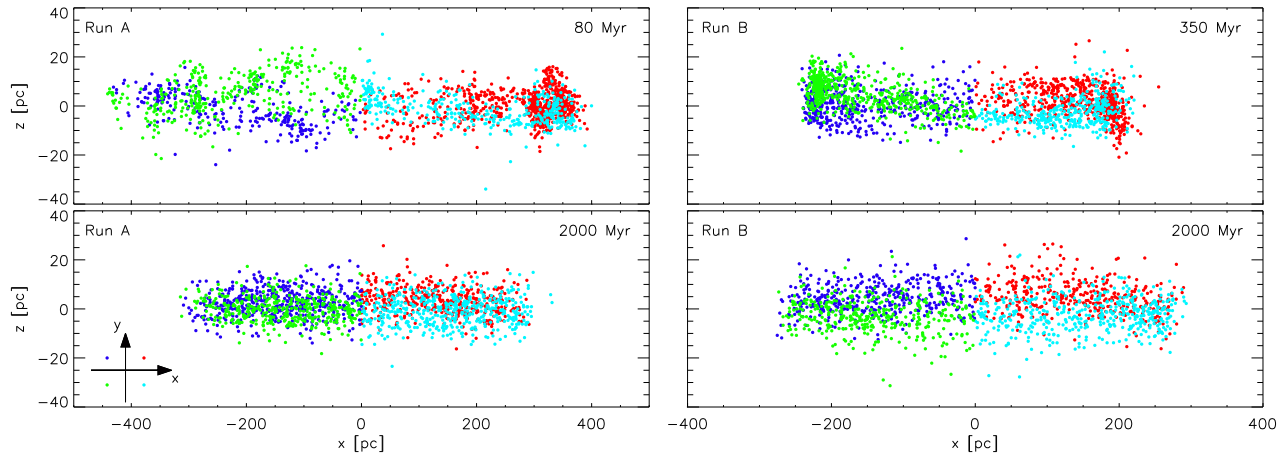}
\caption{Projected distribution of gas particles in runs~A (left panels) and B (right panels). 
For a clear appearance, only a small portion of the gas particles are shown in each panel. 
Blue and red dots represent the gas particles that are located behind the y-axis, while green 
and cyan dots in front of the y-axis.}\label{infinity}
\end{figure*}

\begin{figure}
\centering
\includegraphics{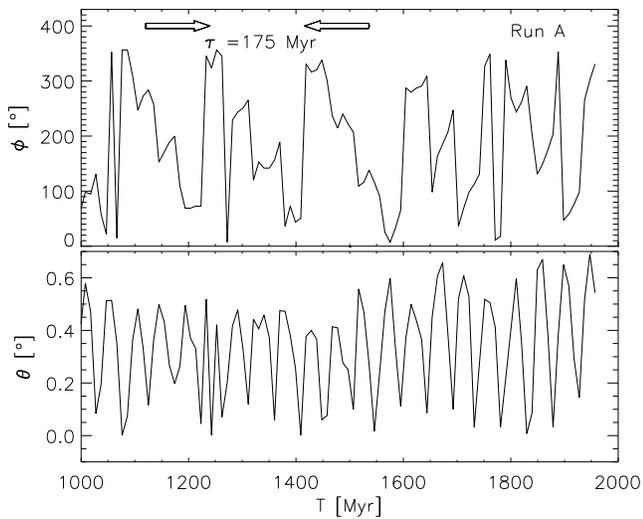}
\caption{Evolution of $\phi$ and $\theta$ of the nuclear ring with respect to
the Galaxy's rotational axis. The tilted nuclear ring precesses along the
Galactic structure with a period of $\sim175~$Myr (upper panel), while the
tilted angle nutates from $\sim0^{\circ}$ to $\sim0.6^{\circ}$ (bottom panel).}
\label{precession}
\end{figure}

The vertical asymmetry of the central mass distribution, although much weaker than 
the asymmetries in the $x$- and $y$-directions, causes the gas clouds infalling to the nuclear 
ring to have vertical components of motion. Through collisions between the gas  clouds orbiting in the 
nuclear ring, the $z$-directional oscillations are likely synchronized to locally have the same 
phase and frequency. Thus, the nuclear ring projected onto the $x$-$z$ (or $y$-$z$) plane exhibits 
a swirling shape. Interestingly, during the first $\sim100$~Myr for type A 
($\sim400$~Myr for type B),  the vertical frequency of the gas motion is twice that of an 
orbital frequency, thus the projected nuclear ring shows a twisted , $\infty$-like shape 
(top panels of Figure \ref{infinity}), which is similarly observed in the CMZ \citep{mol11}. 
However, the $\infty$-like feature does not last for a long time in our simulations, and the vertical oscillations of the whole nuclear ring are synchronized to have the same frequency as that of the orbital motion due to self-interactions of the gas clouds. Since the gas clouds in the nuclear ring follow an X$_2$-type closed orbit, the tilted nuclear ring precesses with the same pattern speed as the Galactic structures (bottom panels of Figure \ref{infinity}).
Figure~\ref{precession} shows that the azimuthal angle of the tilted nuclear ring $\phi$ oscillates 
with a period of $\sim175~$Myr, the rotation period of the Galactic structures. Meanwhile, the 
tilted angle from the Galactic rotational axis $\theta$ is nutating with a period of $\sim50~$Myr.

\begin{figure*}
\centering
\includegraphics{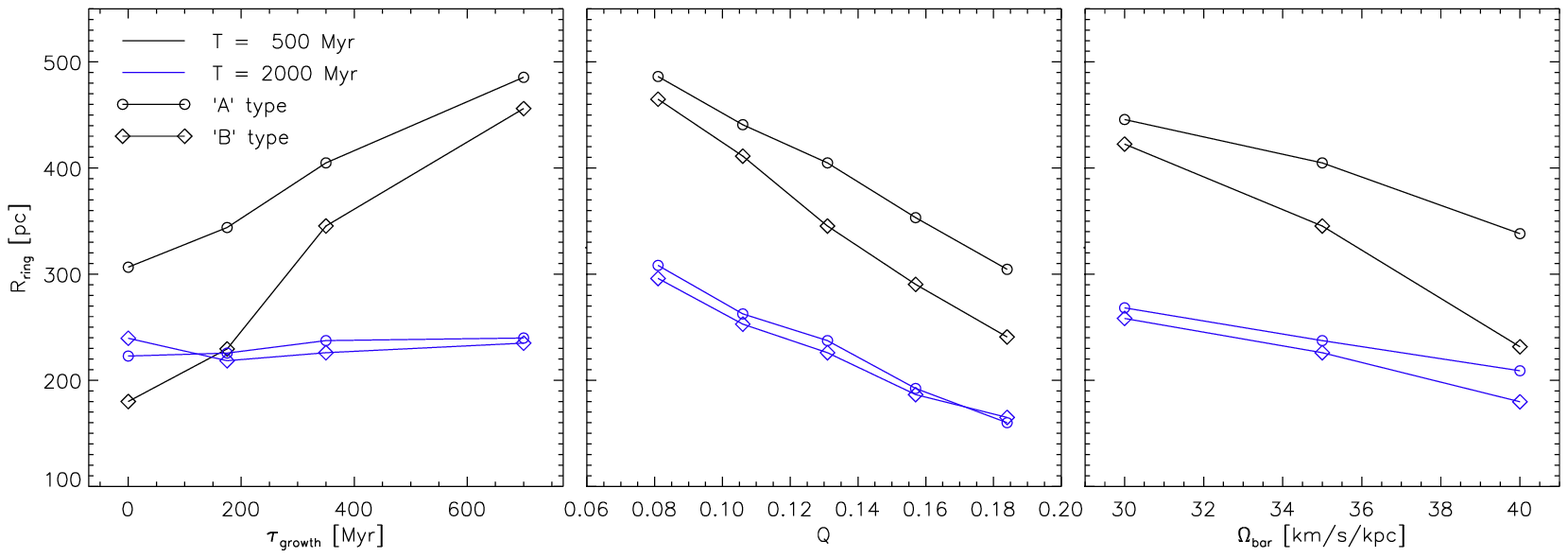}
\caption{Dependence of $R_{ring}$ on $\tau_{growth}$, $Q$, and $\Omega_{bar}$
at $T=500$ and 2000~Myr from our simulations.
The runs of type~A and B are marked with open circles and diamonds, respectively,
at the corresponding locations in each panel:
In the leftmost panel, runs~A(B), At1(Bt1), At2(Bt2), and At3(Bt3)
(where the values of $\tau_{growth}$ are set to 0, 175, 300, and 700~Myr,
respectively; c.f. Table~ref{tbl-3})
are marked with the open symbols from left to right.
Similarly, in the middle panel,
runs~Aq1(Bq1), Aq2(Bq2), A(B), Aq3(Bq3), and Aq4(Bq4)
(with $Q=0.081$, 0.106, 0.131, 0.157, and 0.184, respectively)
are marked.
In the rightmost panel, runs~Ap1(Bp1), A(B), and Ap2(Bp2)
(with $\Omega_{bar}=30$, 35, and 40, respectively)
are shown.}
\vspace{0.5cm}
\label{size}
\end{figure*}

\subsection{Parametric variations}

We examine the dependence of $R_{ring}$ on the parameters $\tau_{growth}$,
$Q$, and $\Omega_{bar}$ (Figure~\ref{size}).
By considering the four runs in each type $-$ A(B) and At1(Bt1) through At3(Bt3) $-$
where the four different periods of $\tau_{growth}$ are adopted,
we find that a longer $\tau_{growth}$ results in a larger $R_{ring}$
at $T=500~$Myr (leftmost panel).
However, since the full $Q$ of these runs are given to be the same,
the increasing tendency of $R_{ring}$ with $\tau_{growth}$
is dissipated at $T=2000~$Myr.
From this, we infer that the current size of the CMZ
does not reflect the early stage of the Galactic bar.

The more strongly barred galaxies have been reported to possess
a smaller-sized nuclear ring \citep{com09,com10,maz11,kim12}.
Likewise, our runs~A(B) and Aq1(Bq1)$-$Aq4(Bq4) show that
$R_{ring}$ decreases with increasing $Q$ (middle panel).
While the standard $Q$ results in the equilibrated nuclear ring of $R_g\sim240~$pc,
a stronger $Q$ causes a smaller-sized nuclear ring of $R_g\sim160-200~$pc,
which better resembles the CMZ. The Galactic structure of B15 is measured to be 
in process of varying $\Omega_{bar}$ from 30 to 40~km/s/kpc for $\sim1~$Gyr,
while the typical $\Omega_{bar}$ is $\sim35~$km/s/kpc.
Our runs~A(B) and At1(Bt1)$-$At2(Bt2) show that
as $\Omega_{bar}$ increases from 30 to 40~km/s/kpc, the equilibrated $R_{ring}$
decreases from $\sim260~$pc to $\sim180~$pc (rightmost panel).

\section{Summary \& Discussion}

In this paper, we performed high-resolution SPH simulations to trace
the formation and evolution of the CMZ under the influence of realistic Galactic structures.
For this, we adopted a snapshot of the self-consistent Galaxy simulation
that reproduces the Galactic stellar disk, grand-design spiral arms, and bar (B15)\footnote{It is 
worth noting that formation/evolution of the realisitic Galactic structures can be described with series 
of snapshots and their ME coefficient sets.}.
The ME method was utilized to efficiently include the realistic Galactic structures
of B15 into our SPH simulations. We found that the Galactic structure is best reproduced 
when: 1) stellar particles are decomposed into nucleus+bulge and disk components
using three exponential disk profiles, 2) the 3D spherical ME model (HO92) is adopted 
for describing the nucleus+bulge component, while the 2D polar ME model (AI78) is used
for the disk component, and 3) the non-zero thickness of the disk component is corrected
by the thick disk correction \citep{bin08}. Our method not only accurately reproduces the 
overall distribution of the stellar particles, but also reduces the calculation time by 
a factor of $\sim30$ compared to that of the full $N$-body calculation of B15.
Furthermore, by modifying the $m=2,4$ ME coefficients, we can easily enlarge/diminish 
the bar elongation and thus imitate the gradual bar growth.

Our SPH simulations reproduce the following overall gaseous structures: 1) the gas 
clouds in the disk region inflow to the $R_g\sim3~$kpc region along the spiral streams, 
2) the gas clouds collide with themselves near the $\mathrm{X}_1$ cusps and plunge into the 
nuclear region along the dust lanes, and 3) the inflowing gas clouds form the nuclear ring. 
Depending on $\tau_{growth}$ and $R_{\mathrm{IB}}$, the size of the nuclear 
rings are different as $R_{ring}\sim150-550~$pc at first. Regardless of the initial $R_{ring}$, they 
converge to $\sim240~$pc at $T\sim1500~$Myr. 

The balanced $R_{ring}$ at $\sim240$~pc is slightly larger than the outer boundary of the CMZ, $\sim200$~pc \citep{mor96} but is significantly larger than the typical radii for the dense gas clouds of the CMZ, $\sim100$~pc \citep{kru15a,hen16a}. The size difference between the nuclear ring (by simulations) and the CMZ (by observations) indicates that our Galactic model lacks an accurate central mass distribution (e.g., \citealt{lau02}).

Recent works by \citet{sor15} and \citet{kru15b} showed that the CMZ resides at much smaller radii than the Galactic ILR, $\sim1$~kpc. Similarly, the nuclear ring in our simulations forms at farther inside than the ILR. The balanced $R_{ring}$ at $\sim240$~pc is quite similar to the stalling radius \citep{kru15b}, which is estimated as $\sim270$~pc in our Galactic model. In this study, we cannot accurately determine the correlation of the balanced $R_{ring}$ with the stalling radius and/or with the angular momentum loss of the inflowing gas clouds at the dust lanes, but it is clear that an $\mathrm{X}_2$ type nuclear ring can settle in farther inside than the ILR.

Interestingly, a very compact nuclear ring is produced when the nuclear structure of $R_g<1$~kpc is initially removed; thus, the inflowing gas clouds do not collide with the nuclear ring structure. The inflowing gas clouds to the nuclear region form a very compact nuclear ring of $R_g\sim150-200$~pc during the first few hundred Myr. We have hypothesized that a small merging galaxy approaching close to the galactic center might disperse the existing nuclear structure and perturb the gaseous disk \citep{lan13}. In a follow-up study, we aim to study how a minor merging event affects the redistribution of gaseous nuclear structures.

As the gas clouds in the disk are depleted after $T\sim1200~$Myr, the gas inflow rate to the nuclear ring is equilibrated at $\dot{M}_{gas}\sim0.02\msun/\mathrm{yr}$, which is insufficient to maintain the SFR in the nuclear ring at $\sim0.1\msun/\mathrm{yr}$. Thus, additional mechanisms that can continuously feed the gas reservoirs, such as gas replenishment from stellar mass loss, galactic fountains, and cosmic accretion \citep{seg16,opp08,opp10,ker09,dek09,ric12}, are necessary to trace the realistic $\dot{M}_{gas}$ gas to the central region over a few Gyr timescale. This issue will be discussed in a follow-up study. 

Unlike the low SFR of the CMZ compared to predicted rates \citep{lon13}, the nuclear ring in our simulations overproduces stars relative to the SFR-gas density relation of the CMZ. This occurs because our star formation model is implemented to reproduce the Schmidt law (Equation 8). Meanwhile, the episodic star formation activity whose current stage is at a minimum is suggested to explain the low SFR of the CMZ \citep{kru14,kru15b,kru17}. However, in our simulations, stars in the nuclear ring form consistently rather than in a burst-like manner (see the middle panel of Fig. \ref{inflow}). This difference might be due to the limits of our numerical model that: 1) the minimum size of the SPH kernel is set to smooth high-density clumpy structures, 2) the standard SPH with density-entropy formulation used in our simulations does not adequately address dynamical instabilities and contact discontinuities \citep{age07}, and 3) our feedback models do not reflect radiation feedback from stars, such as position-dependent photoionization and radiation pressure \citep{hop14}.

Thanks to our Galaxy model, which is found to have a lopsided mass distribution, the nuclear ring reproduced by our simulations consistently shows: 1) periodic variation in the asymmetric mass distribution, 2) a twisted, $\infty$-like shape, and 3) tilt with respect to the Galactic plane. The effect of a lopsided mass distribution on the CMZ should be revisited using a more realistic central mass distribution of the Galaxy (e.g., \citealt{lau02,kru15a}).

\acknowledgements
We appreciate the anonymous referee for his/her helpful comments that greatly improved our manuscript. This work was supported by the National Research Foundation grant funded by the Ministry of Science, ICT and Future Planning of Korea (NRF-2014R1A2A1A11052367). JB was supported by HPCI Strategic Program Field 5 `The origin of matter 
and the universe' and JSPS Grant-in-Aid for Young Scientists (B) Grant 
Number 26800099. TRS was supported by HPCI Strategic Program Field 5 
`The origin of matter and the universe' and JSPS Grant-in-Aid for Young Scientists (A) 
Grant Number 26707007. JSH was supported by Basic Science Research Program 
through the National Research Foundation of Korea (NRF) funded by the Ministry of 
Education (Grant No. 2015R1D1A1A01059148) and by the Korea Astronomy and Space 
Science Institute under the R\&D program (Project No. 2015-1-320-18) supervised 
by the Ministry of Science, ICT, and Future Planning. KC was supported by the 
BK21 plus program through the NRF funded by the Ministry of Education of Korea, 
and by the Supercomputing Center/Korea Institute of Science and Technology 
Information with supercomputing resources including technical support (KSC-2014-C2-005).

\appendix

\section{A. 3D spherical multipole expansion model of Hernquist \& Ostriker (1992)}

For a computational convenience, density $\rho$ and potential $\Phi$ in a position of $(r,\theta,\phi)$ are rewritten as
\begin{eqnarray}
\rho(r,\theta,\phi)=\sum_{l=0}^{\infty}\sum_{m=0}^{l}P_{lm}(\cos~\theta)[A_{lm}(r)\cos~m\phi+B_{lm}(r)\sin~m\phi],\nonumber\\
\Phi(r,\theta,\phi)=\sum_{l=0}^{\infty}\sum_{m=0}^{l}P_{lm}(\cos~\theta)[C_{lm}(r)\cos~m\phi+D_{lm}(r)\sin~m\phi],
\end{eqnarray}
where $P_{lm}$ is a Legendre function, and $A_{lm}$, $B_{lm}$, $C_{lm}$, and $D_{lm}$ are
\begin{equation}
\left[ \begin{array}{c}
  A_{lm}(r) \\
  B_{lm}(r) \\
  C_{lm}(r) \\
  D_{lm}(r)
  \end{array}\right] = N_{lm}\sum_{n=0}^{\infty}\widetilde{A}_{nl}\left[ \begin{array}{c}
  \widetilde{\rho}_{nl}(r)\\
  \widetilde{\rho}_{nl}(r)\\
  \widetilde{\Phi}_{nl}(r)\\
  \widetilde{\Phi}_{nl}(r)
  \end{array}\right] \left[\begin{array}{c}
  \Sigma_{nlm}(\cos)\\
  \Sigma_{nlm}(\sin)\\
  \Sigma_{nlm}(\cos)\\
  \Sigma_{nlm}(\sin)
  \end{array}\right].
\end{equation}
Here, $N_{lm}$, $\widetilde{A}_{nl}$, $\widetilde{\rho}_{nl}$, and $\widetilde{\Phi}_{nl}$ are
\begin{eqnarray}
N_{lm}&=&\frac{2l+1}{4\pi}(2-\delta_{m0})\frac{(l-m)!}{(l+m)!},\nonumber\\
\widetilde{A}_{nl}&=&-\frac{2^{8l+6}}{K_{nl}~4\pi}\frac{n!~(n+2l+3/2)[\Gamma(2l+3/2)]^2}{\Gamma(n+4l+3)},\nonumber\\
\widetilde{\rho}_{nl}&=&\frac{K_{nl}}{2\pi}\frac{r^l}{r(1+r)^{2l+3}}C_{n}^{(2l+3/2)}\left(\frac{r-1}{r+1}\right)\sqrt{4\pi},\nonumber\\
\widetilde{\Phi}_{nl}&=&-\frac{r^l}{(1+r)^{2l+1}}C_{n}^{(2l+3/2)}\left(\frac{r-1}{r+1}\right)\sqrt{4\pi},
\end{eqnarray}
where $C_n^{\alpha}$ is a Gegenbauer polynomial, and $K_{nl}$ is a normalization constant of
\begin{equation}
K_{nl}=n(n+4l+3)/2+(l+1)(2l+1).
\end{equation}
The expansion coefficients of $\Sigma_{nlm}(\cos)$ and $\Sigma_{nlm}(\sin)$ are calculated with a collection of $k$th particles by,
\begin{equation}
\left[\begin{array}{c}
    \Sigma_{nlm}(\cos)\\
    \Sigma_{nlm}(\sin)
    \end{array}\right]=\sum_{k}m_{k}\widetilde{\Phi}_{nl}(r_k)P_{lm}(\cos~\theta_k)\left[ \begin{array}{c}
    \cos~m\phi_k\\
    \sin~m\phi_k
  \end{array}\right].
\end{equation}

Similar to $\rho$ and $\Phi$, accelerations $a_r$, $a_{\theta}$, and $a_{\phi}$ are derived by
\begin{eqnarray}
a_r(r,\theta,\phi)&=&-\sum_{l=0}^{\infty}\sum_{m=0}^{l}P_{lm}(\cos~\theta)[E_{lm}(r)\cos~m\phi+F_{lm}(r)\sin~m\phi],\nonumber\\
a_{\theta}(r,\theta,\phi)&=&-\frac{1}{r}\sum_{l=0}^{\infty}\sum_{m=0}^{l}\frac{dP_{lm}(\cos~\theta)}{d\theta}[C_{lm}(r)\cos~m\phi+D_{lm}(r)\sin~m\phi],\nonumber\\
a_{\phi}(r,\theta,\phi)&=&-\frac{1}{r}\sum_{l=0}^{\infty}\sum_{m=0}^{l}\frac{mP_{lm}(\cos~\theta)}{\sin~\theta}[D_{lm}(r)\cos~m\phi-C_{lm}(r)\sin~m\phi],
\end{eqnarray}
where $E_{lm}$ and $F_{lm}$ are
\begin{equation}
\left[\begin{array}{c}
    E_{lm}(r)\\
    F_{lm}(r)
    \end{array}\right]=N_{lm}\sum_{n=0}^{\infty}\widetilde{A}_{nl}\frac{d}{dr}\widetilde{\Phi}_{nl}(r)\left[\begin{array}{c}
    \Sigma_{nlm}(\cos)\\
    \Sigma_{nlm}(\sin)
    \end{array}\right].
\end{equation}

\section{B. 2D disk multipole expansion model of Aoki \& Iye (1978)}
Surface density $\mu$ and potential $\Psi$ in a cylindrical coordinate for the disk system are given by
\begin{eqnarray}
\mu(R,\phi)&=&\sum_{n=0}^{\infty}\sum_{m=0}^{\infty}\left[A_{nm}(R)\cos~m\phi+B_{nm}(R)\sin~m\phi\right],\nonumber\\
\Psi(R,\phi)&=&\sum_{n=0}^{\infty}\sum_{m=0}^{\infty}\left[C_{nm}(R)\cos~m\phi+D_{nm}(R)\sin~m\phi\right],
\end{eqnarray}
where
\begin{equation}
\left[ \begin{array}{c}
  A_{nm}(R) \\
  B_{nm}(R) \\
  C_{nm}(R) \\
  D_{nm}(R)
  \end{array}\right] = N_{nm}\left[ \begin{array}{c}
  \widetilde{\mu}_{nm}(R)\\
  \widetilde{\mu}_{nm}(R)\\
  \widetilde{\Psi}_{nm}(R)\\
  \widetilde{\Psi}_{nm}(R)
  \end{array}\right]\left[\begin{array}{c}
  \Sigma_{nm}(\cos)\\
  \Sigma_{nm}(\sin)\\
  \Sigma_{nm}(\cos)\\
  \Sigma_{nm}(\sin)
  \end{array}\right].
\end{equation}
Here, $N_{nm}$, $\widetilde{\mu}_{nm}$, and $\widetilde{\Psi}_{nm}$ are
\begin{eqnarray}
N_{nm}&=&-2(2-\delta_{m0})\frac{(n-m)!}{(n+m)!},\nonumber\\
\widetilde{\mu}_{nm}&=&=\frac{2n+1}{2\pi}\frac{M}{a^2}{\left(\frac{a^2}{r^2+a^2}\right)}^{3/2}P_{nm}\left(\frac{r^2-a^2}{r^2+a^2}\right),\nonumber\\
\widetilde{\Psi}_{nm}&=&=-\frac{GM}{a}{\left(\frac{a^2}{r^2+a^2}\right)}^{1/2}P_{nm}\left(\frac{r^2-a^2}{r^2+a^2}\right).
\end{eqnarray}
The expansion coefficients $\Sigma_{nm}(\cos)$ and $\Sigma_{nm}(\sin)$ can be calculated with a collection of $k$th particles by,
\begin{equation}
\left[\begin{array}{c}
    \Sigma_{nm}(\cos)\\
    \Sigma_{nm}(\sin)
    \end{array}\right]=\sum_{k}m_{k}\widetilde{\Psi}_{nm}(r_k)\left[ \begin{array}{c}
    \cos~m\phi_k\\
    \sin~m\phi_k
  \end{array}\right].
\end{equation}
Similar to $\mu$ and $\Psi$, acceleration $a_R$ and $a_{\theta}$ can be expressed as
\begin{eqnarray}
a_R(R,\phi)&=&-\sum_{n=0}^{\infty}\sum_{m=0}^{\infty}\left[E_{nm}(R)\cos~m\phi+F_{nm}(R)\sin~m\phi\right],\nonumber\\
a_{\phi}(R,\phi)&=&-\frac{m}{R}\sum_{n=0}^{\infty}\sum_{m=0}^{\infty}\left[D_{nm}(R)\cos~m\phi-C_{nm}(R)\sin~m\phi\right],
\end{eqnarray}
where $E_{nm}$ and $F_{nm}$ are
\begin{equation}
\left[\begin{array}{c}
    E_{nm}(r)\\
    F_{nm}(r)
    \end{array}\right]=N_{nm}\frac{d}{dR}\widetilde{\Psi}_{nm}(R)\left[\begin{array}{c}
    \Sigma_{nm}(\cos)\\
    \Sigma_{nm}(\sin)
    \end{array}\right].
\end{equation}



\end{document}